\documentclass[preprintnumbers,article,amsmath,amssymb,floatfix,10pt,prd,onecolumn,superscriptaddress,nofootinbib]{revtex4-2}

\usepackage{titlesec}
\usepackage[toc]{appendix}
\usepackage{caption}
\titleformat*{\section}{\LARGE\bfseries}
\titleformat*{\subsection}{\Large\bfseries}
\titleformat*{\subsubsection}{\large\bfseries}
\titleformat*{\paragraph}{\large\bfseries}
\titleformat*{\subparagraph}{\large\bfseries}
\usepackage{bm}
\usepackage{amsfonts}
\usepackage{latexsym}
\usepackage[utf8]{inputenc}
\usepackage{graphicx}
\usepackage{amsmath}
\usepackage{palatino}
\usepackage{mathpazo}
\usepackage{textcomp}
\usepackage{float}
\usepackage{booktabs}
\usepackage{dcolumn}
\usepackage{ragged2e}
\usepackage{hyperref}
\hypersetup{colorlinks,citecolor=blue}
\hypersetup{colorlinks=true,linkcolor=blue,filecolor=magenta,urlcolor=blue}
\usepackage{amsthm}
\usepackage{xcolor}
\usepackage{orcidlink}
\usepackage{epsfig}
\usepackage{commath}
\usepackage{cancel, ulem}
\usepackage{subcaption}
\usepackage{caption}
\usepackage{multirow}
\newcommand{\m}{\mathring}
\newcommand{\be}{\begin{equation}}
\newcommand{\ee}{\end{equation}}
\newcommand{\bea}{\begin{eqnarray}}
\newcommand{\eea}{\end{eqnarray}}
\newcommand{\eeas}{\end{eqnarray*}}
\newcommand{\beas}{\begin{eqnarray*}}

\def\jnl@style{\it}
\def\aaref@jnl#1{{\jnl@style#1}}

\def\aaref@jnl#1{{\jnl@style#1}}

\def\aj{\aaref@jnl{AJ}}                   
\def\apj{\aaref@jnl{ApJ}}                 
\def\apjl{\aaref@jnl{ApJ}}                
\def\apjs{\aaref@jnl{ApJS}}               
\def\apss{\aaref@jnl{Ap\&SS}}             
\def\aap{\aaref@jnl{A\&A}}                
\def\aapr{\aaref@jnl{A\&A~Rev.}}          
\def\aaps{\aaref@jnl{A\&AS}}              
\def\mnras{\aaref@jnl{Mon.~Not.~Roy.~Astron.~Soc.}}             
\def\prd{\aaref@jnl{Phys.~Rev.~D}}        
\def\prc{\aaref@jnl{Phys.~Rev.~C}}  
\def\prl{\aaref@jnl{Phys.~Rev.~Lett.}}    
\def\qjras{\aaref@jnl{QJRAS}}             
\def\skytel{\aaref@jnl{S\&T}}             
\def\ssr{\aaref@jnl{Space~Sci.~Rev.}}     
\def\zap{\aaref@jnl{ZAp}}                 
\def\nat{\aaref@jnl{Nature}}              
\def\aplett{\aaref@jnl{Astrophys.~Lett.}} 
\def\apspr{\aaref@jnl{Astrophys.~Space~Phys.~Res.}} 
\def\physrep{\aaref@jnl{Phys.~Rep.}}      
\def\physscr{\aaref@jnl{Phys.~Scr}}       
\def\commat{\aaref@jnl{Comm.~Math.~Phys.}}              
\def\science{\aaref@jnl{Science}}               
\def\cqg{\aaref@jnl{Classical Quant.~Grav.}}            
\def\jpcs{\aaref@jnl{JPCS}}                                     
\def\ijmpd{\aaref@jnl{Int.~J.~Mod.~Phys.~D}}                    
\def\grg{\aaref@jnl{Gen.~Relat.~Gravit.}}               
\def\rpp{\aaref@jnl{Rep.~Prog.~Phys.}}          
\def\npa{\aaref@jnl{Nucl.~Phys.~A}}        
\def\lrr{\aaref@jnl{Living Rev.~Rel.}}                   
\def\jcap{\aaref@jnl{J.~Cosmology Astropart.~Phys.}}    
\def\rmp{\aaref@jnl{Rev.~Mod.~Phys.}}   
\def\epjc{\aaref@jnl{Eur.~Phys.~J.~C}} 
\def\plb{\aaref@jnl{~Phy.~Lett.~B}} 
\def\mpla{\aaref@jnl{Mod.~Phy.~Lett.~A}} 
\def\arxiv{\aaref@jnl{arxiv.org}}

\allowdisplaybreaks[1]

\begin{document}

\title{Role of the Dynamic Degree of Freedom in Scalar-Tensor Non-Metricity Gravity with Curved FLRW Geometry}
\author{Ghulam Murtaza\orcidlink{0009-0002-6086-7346}}
\email{ghulammurtaza@1utar.my}
\affiliation{Department of Mathematical and Actuarial Sciences, Universiti Tunku Abdul Rahman, Jalan Sungai Long,
43000 Cheras, Malaysia}
\author{Avik De\orcidlink{0000-0001-6475-3085}}
\email{avikde@um.edu.my}
\affiliation{Institute of Mathematical Sciences, Faculty of Science, Universiti Malaya, 50603 Kuala Lumpur, Malaysia}
\author{Andronikos Paliathanasis\orcidlink{0000-0002-9966-5517}}
\email{anpaliat@phys.uoa.gr}
\affiliation{Department of Mathematics, Faculty of Applied
Sciences, Durban University of Technology, Durban 4000, South Africa}
\affiliation{School for Data Science and Computational Thinking, Stellenbosch 
University,44 Banghoek Rd, Stellenbosch 7600, South Africa}
\affiliation{Departamento de Matem\`{a}ticas, Universidad Cat\`{o}lica del 
Norte, Avda. Angamos 0610, Casilla 1280 Antofagasta, Chile}
\affiliation{National Institute for Theoretical and Computational Sciences (NITheCS), South Africa}

\footnotetext{The research has been carried out under Universiti Tunku Abdul Rahman Research Fund project IPSR/RMC/UTARRF/2023-C1/A09 provided by Universiti Tunku Abdul Rahman. }

\begin{abstract}
We investigate a non-minimally coupled scalar field theory within the framework of scalar-tensor non-metricity gravity, focusing on spatially curved FLRW spacetimes. Employing the dynamical systems approach with Hubble-normalized variables, we reformulate the field equations into an autonomous system and analyze the resulting critical points. Four distinct cases, determined by the scalar coupling and potential functions, are studied in detail. For each case, we identify the existence and stability of equilibrium points, classify their cosmological behavior, and compute key observables such as the deceleration parameter and effective equation of state. Our results reveal that the theory admits matter-dominated eras, parameter-dependent saddle solutions, and stable de Sitter attractors capable of driving late-time cosmic acceleration. The additional scalar degree of freedom introduced by the non-coincident gauge plays a crucial role in determining the system’s dynamics and viability. These findings emphasize the potential of scalar-tensor non-metricity gravity as a robust extension of general relativity and motivate further confrontation of the model with observational data.

\end{abstract}
\maketitle

\section{Introduction}\label{sec0}
Recent cosmological observations \cite{Riess, Tegmark, Kowalski,Komatsu,Abdalla,Lynch2024} have established that the General Theory of Relativity (GR) cannot fully explain the evolution of the universe and exhibits several shortcomings. These limitations have motivated researchers to explore theories beyond GR to uncover the hidden mysteries of the cosmos. As a result, considerable effort has been devoted by the scientific community to develop modified theories of gravity, which serve as alternatives or extensions to GR \cite{Clifton2012}. The simplest extension of GR is the $f(\mathring{R})$ theory, in which the Ricci scalar $\mathring{R}$ in the action term is replaced by an arbitrary function of itself \cite{Felice2010}. Researchers further attempted to modify the theory of gravity using geometrical instruments. By departing from the Levi-Civita construction, where the connection is torsion-free and metric-compatible, and allowing for non-vanishing torsion, one naturally arrives at the metric teleparallel formulation of gravity. In this framework, a notable extension is the $f(\mathbb{T})$ theory, which is based on a connection with vanishing curvature, non-zero torsion, and zero non-metricity \cite{baha23023}. Another alternative is the symmetric teleparallel formulation, characterized by a connection with vanishing curvature and torsion, but non-vanishing non-metricity \cite{heisen2024}. Within the Friedmann-Lemaître-Robertson-Walker (FLRW) cosmological framework, the symmetric teleparallel approach admits four distinct classes of affine connections compatible with its symmetries: three in the spatially flat case; say, denoted by $\Gamma_1$, $\Gamma_2$ and $\Gamma_3$, and one for non-zero spatially curved case, denoted by $\Gamma_4$. This comes with an interesting aspect that connection $\Gamma_4$ can be seamlessly reduced to $\Gamma_3$ in the spatially flat limit  \cite{FLRW/connection}. The defining feature of symmetric teleparallel gravity ensures the existence of a special coordinate system in which the affine connection vanishes identically and the covariant derivative reduces to the ordinary partial derivative. As a result, inertial effects can be clearly disentangled from genuine gravitational dynamics. Connection related to this choice of coordinates is therefore referred to as coincident gauge connection, while all others are termed as non-coincident gauges. The choice of connection plays a crucial role in the formulation of the theory. In fact, selecting a specific connection as the starting point leads to dynamically in-equivalent gravitational theories, even within the same symmetric teleparallel framework. Notably, in the coincident gauge, the background field equations of symmetric teleparallel gravity reduce to those of metric teleparallel equivalent gravity, implying that no genuinely new cosmological dynamics emerge at the background level in this case. To uncover novel phenomenology intrinsic to symmetric teleparallel gravity and to fully explore its dynamical richness, it is therefore necessary to go beyond the coincident gauge.
\\\\

Scalar-tensor gravity has a long-established role in the curvature-based formulation of General Relativity, where the gravitational dynamics are encoded in the Ricci scalar $\mathring{R}$. A broad and widely studied class of curvature-based scalar-tensor theories is described by the action \cite{Far2004,Rattazzi2009}
\begin{align}
S_{(\mathring{R})}=\int d^4x\,\sqrt{-g}\left[
\frac{1}{2}\,f(\phi)\,\mathring{R}
-\frac{1}{2}\,h(\phi)\,g^{\mu\nu}\partial_\mu\phi\,\partial_\nu\phi
-U(\phi)
\right],
\end{align}
which includes, as particular cases, Brans-Dicke-type models and many of their generalizations \cite{Brans1961}.  $U(\phi)$ is the scalar field potential, $f(\phi)$ is the coupling function. These theories are strongly motivated by both cosmological and fundamental considerations, since the scalar field can act as an effective varying gravitational coupling, drive inflationary dynamics, or support late-time cosmic acceleration.\\

In parallel, within the torsion-based teleparallel formulation of gravity (metric teleparallel gravity), where the gravitational interaction is encoded in the torsion scalar $\mathbb T$, an analogous extension is given by scalar-torsion gravity. In this framework, a representative non-minimally coupled scalar-torsion theory can be written as \cite{Poplawski2020,Skugoreva2015}
\begin{equation}
S_{(\mathbb T)}=\int d^4x\,\sqrt{-g}\left[
\frac{1}{2}\,f(\phi)\,\mathbb T
-\frac{1}{2}\,h(\phi)\,g^{\mu\nu}\partial_\mu\phi\,\partial_\nu\phi
-U(\phi)
\right].
\end{equation}
This shows that the non-minimal scalar coupling programme can also be consistently implemented in a torsional representation of gravity.\\

Motivated by these two parallel developments, it is natural to investigate the corresponding scalar-tensor extension within the third and complementary geometrical formulation of gravity, namely symmetric teleparallel gravity, where the gravitational interaction is described by the non-metricity scalar $Q$. This leads to the non-metricity analogue of scalar-tensor theory introduced in \cite{Saal2018}, which can be expressed as
\begin{equation}
S_{(Q)}=\int d^4x\,\sqrt{-g}\left[
\frac{1}{2}\,f(\phi)\,Q
-\frac{1}{2}\,h(\phi)\,g^{\mu\nu}\partial_\mu\phi\,\partial_\nu\phi
-U(\phi)
\right].
\end{equation}
Therefore, our main motivation is that scalar-tensor gravity has already been extensively studied in both the curvature-based and torsion-based formulations, and it is timely and well-motivated to construct and explore the corresponding non-metricity-based scalar-tensor extension. In this way, the scalar-non-metricity coupling completes the geometric trinity of scalar-tensor theories of gravity based on curvature, torsion, and non-metricity. In another way, we can say that the scalar fields which were once central in the early universe through the inflaton driving inflation, have re-emerged as important even in late-time cosmology.\\

A considerable amount of research has focused on achieving a unified description of the universe by introducing a single component that mimics dust-like matter during the early and intermediate epochs, while acting as the driver of cosmic acceleration at late times. Such a unification of the matter-dominated and dark energy eras can be realized not through exotic or ad-hoc fluids, but within a broader class of scalar-torsion theories investigated in \cite{Basilakos2022}. The scalar-tensor gravity in the non-metricity context has been first explored in \cite{Saal2018}, and the cosmological aspects have been explored. In a subsequent paper \cite{laur2024}, the authors have investigated the alternative FLRW connections, which introduce an extra degree of freedom that significantly modifies scalar field dynamics but cannot mimic dark matter or dark energy. They have further shown that the stability of the standard cosmological eras is possible under certain restrictions, although finite-time singularities may also arise. A Brans-Dicke theory within non-metricity gravity was investigated in \cite{BD024}, where the exact cosmological solutions exhibit invariant physical properties under conformal transformation between Jordan and Einstein frames, and the first analytic solution for symmetric teleparallel scalar-tensor cosmology was also provided. In \cite{Giacomini2024}, it was shown that the equilibrium points and accelerated cosmological solutions are preserved under conformal transformations, establishing a one-to-one correspondence between the Jordan and Einstein frames in scalar non-metricity gravity. In \cite{murtazascalar}, an interesting comparison of scalar non-metricity and scalar-torsion theories has been presented in a spatially-flat FLRW model. In \cite{Valcarcel2022}, exact scalarised spherical solutions in the framework of non-metricity scalar-tensor  gravity have been presented.\\

In cosmology, dynamical system analysis (DSA) has proven to be a powerful mathematical tool that reformulates the field equations using dimensionless variables, resulting in a coupled system of first-order algebraic differential equations. Within this dynamical system, we identify and analyze the stationary (critical) points, each corresponding to a distinct phase in the cosmological evolution. Furthermore, we perform a stability analysis of these points, which is essential for evaluating the physical viability and consistency of the underlying cosmological model across different epochs of the universe. For more details on significant DSA works in a varied range of modified and scalar tensor theories, see \cite{Bahamonde2018b, Wainwright1997, Odintsov2017a,fQBI, Murtaza2025, Carloni:2024ybx, saridakisST} and the references therein. \\

In most cosmological studies, the observable universe is typically assumed to be exactly spatially flat. However, this assumption should ideally be re-evaluated and constrained each time new observational datasets become available. Consequently, it is important to consider the role of spatial curvature $k$ in cosmological analyses. Recent investigations \cite{Giare2023, Giacomini2022,Cruz2018, Melchiorri2020,Vagnozzi2021,Loeb2021, Alsing2021,Lai2023, Subram2023,Jen2025} have focused on the impact of non-zero curvature, highlighting its significance. In this context, it becomes particularly relevant to explore new avenues within curved FLRW geometry. In addition, recently, curved inflationary models have been explored, showing that inflation is not affected by negative curvature, while the curvature energy density can remain non-zero in the pre-inflationary stage, and through the cosmological principle, one obtains homogeneous and isotropic open or closed scenarios that asymptotically evolve toward spatial flatness at late times \cite{Steigman1998,Mathews2015,Aslanyan2015}. In literature, a few DSA works by considering non-flat spacetime have been studied \cite{Luongo2025,Dunsby,Kerachian20199}. For instance, in the context of $f(Q)$ theory \cite{Loo2024}, it was demonstrated that curvature generates new critical points, including inflationary, dark matter, and dark energy solutions, offering a possible resolution to the coincidence problem and cosmological tensions. Similarly, in \cite{Paliathanasis:2023raj}, nonlinear models naturally admit de Sitter solutions as unique attractors, allowing small deviations from Symmetric Teleparallel General Relativity (STGR) to address the flatness problem without a cosmological constant.\\

Motivated, in this study we consider the scalar non-metricity gravity and investigate the dynamical system analysis of spatially curved as well as flat spacetime in a unified manner by employing the connection classes $\Gamma_4$ and $\Gamma_3$, respectively. This approach allows the yet-to-be-derived field equations to be reformulated as an equivalent system of algebraic differential equations, facilitating the identification of fixed points and a detailed analysis of the physical nature of their associated asymptotic solutions. To note that, the second connection branch in spatially flat FLRW spacetime has already been studied in \cite{murtazascalar}, and detailed comparisons with the present framework have been provided.\\

This paper is organized in the following way: after the Introduction in Section \ref{sec0}, we provide a brief overview of the mathematical foundations of symmetric teleparallel theory, followed by the field equations for the non-metricity approach of the scalar tensor gravity in Section \ref{secnew}. In Section \ref{sec1}, we explore the cosmological implications of this theory for the compatible connection classes. A comprehensive dynamical system analysis of this theory in a unified manner in both spatially flat and non-flat universes, considering varied choices of the coupling function and potential, is presented in Section \ref{sec2} and its subsections. Finally, our main results and conclusions are summarized in Section \ref{sec3}.

\section{Non-metricity version of scalar-tensor gravity theory} \label{secnew}
The Levi-Civita connection $\mathring{\Gamma}^\alpha{}_{\mu\nu}$ is the unique affine connection with the combined property of metric-compatibility and torsion-free and thus it can be presented in terms of the metric $g$
\begin{equation}
\mathring{\Gamma}^\alpha_{\,\,\,\mu\nu}=\frac{1}{2}g^{\alpha\beta}\left(\partial_\nu g_{\beta\mu}+\partial_\mu g_{\beta\nu}-\partial_\beta g_{\mu\nu}  \right)\,,
\end{equation}
However, we can always consider a torsion-free and curvature-free affine connection $\Gamma^\alpha{}_{\mu\nu}$, with the property of non-vanishing non-metricity tensor
\begin{equation} \label{Q tensor}
Q_{\lambda\mu\nu} := \nabla_\lambda g_{\mu\nu}=\partial_\lambda g_{\mu\nu}-\Gamma^{\beta}_{\,\,\,\lambda\mu}g_{\beta\nu}-\Gamma^{\beta}_{\,\,\,\lambda\nu}g_{\beta\mu}\neq 0 \,.
\end{equation}
We present
\begin{equation} \label{connc}
\Gamma^\lambda{}_{\mu\nu} := \mathring{\Gamma}^\lambda{}_{\mu\nu}+L^\lambda{}_{\mu\nu}
\end{equation}
where $L^\lambda{}_{\mu\nu}$ is the disformation tensor, given by
\begin{equation} \label{L}
L^\lambda{}_{\mu\nu} = \frac{1}{2} (Q^\lambda{}_{\mu\nu} - Q_\mu{}^\lambda{}_\nu - Q_\nu{}^\lambda{}_\mu) \,.
\end{equation}

The superpotential (or the non-metricity conjugate) tensor $P^\lambda{}_{\mu\nu}$ is given by
\begin{equation} \label{P}
P^\lambda{}_{\mu\nu} = 
\frac{1}{4} \left( -2 L^\lambda{}_{\mu\nu} + Q^\lambda g_{\mu\nu} - \tilde{Q}^\lambda g_{\mu\nu} -\delta^\lambda{}_{(\mu} Q_{\nu)} \right) \,,
\end{equation}
where
\begin{equation*}
 Q_\mu := g^{\nu\lambda}Q_{\mu\nu\lambda} = Q_\mu{}^\nu{}_\nu \,, \qquad \tilde{Q}_\mu := g^{\nu\lambda}Q_{\nu\mu\lambda} = Q_{\nu\mu}{}^\nu \,.
\end{equation*}
Finally, the non-metricity scalar $Q$ is defined as
\begin{equation} \label{Q}
Q=Q_{\alpha\beta\gamma}P^{\alpha\beta\gamma}\,.
\end{equation}

Mimicking the gravitational action of the scalar-tensor extension of GR, an action was considered in \cite{Saal2018}
\begin{align}\label{eqn:ST}
S=\frac{1}{2\kappa }\int\sqrt{-g}\left[f(\phi)Q-h(\phi)\nabla^\alpha\phi\nabla_\alpha\phi-U(\phi)
+2\kappa\mathcal L_m \right] \,d^{4}x\,.
\end{align} 

At this point, we remark that, by considering the scalar non-metricity theory with $f(\phi)=\phi$ and  $h(\phi)=\frac{\omega}{\phi}$ with $\omega$ = constant, the resulting theory is analogue to the symmetric teleparallel Brans-Dicke theory, where $\omega$ plays the role of the Brans-Dicke parameter. As well-known in scalar-tensor theory, the action is invariant under scalar field reparametrization, which can reduce one of the functions to be a constant. So, without loss of generality, let us redefine the scalar fields to make $h(\phi)$ to be a constant.  It is also important to observe that by setting $f(\phi)=\phi$, $h(\phi)=0$, where now $\phi=f'(Q)$ and $U(\phi)=(f'(Q)Q-f(Q))$ which means that the Action (\ref{eqn:ST}) is equivalent to $f(Q)$ theory. 

The variation of the action term with respect to the metric produces the metric field equations
\begin{align}
\kappa T_{\mu\nu}
=&f\m G_{\mu\nu} +
2f'P^\lambda{}_{\mu\nu}\nabla_\lambda \phi
-h\nabla_\mu\Phi\nabla_\nu\phi
+\frac12hg_{\mu\nu}\nabla^\alpha\phi\nabla_\alpha\phi+\frac12Ug_{\mu\nu} \,,
\label{eqn:FE1}
\end{align}
where 
$\m G_{\mu\nu}$ denotes the Einstein tensor corresponding to the Levi-Civita connection;
$T_{\mu\nu}$ is the stress energy tensor defined as 
\begin{align*}
T_{\mu\nu}=-\frac 2{\sqrt{-g}}\frac{\delta(\sqrt{-g}\mathcal L_M)}{\delta g^{\mu\nu}}\,,
\end{align*}
and 
(~)' means the derivative of the function (~) with  respect to $\phi$.
On the other hand, the variation of the action with respect to the scalar field $\phi$ leads us to the second field equations  
\begin{align}\label{eqn:FE2}
f'Q+h'\nabla^\alpha\phi\nabla_\alpha\phi+2h\m\nabla^\alpha\m\nabla_\alpha \phi-U'=0.
\end{align}
Apart from the metric tensor and the scalar field $\phi$, there is another set of dynamic variables: the components of the affine connection, which yields the connection field equations 
\begin{align}\label{eqn:FE3}
(\nabla_\mu-\tilde L_\mu)(\nabla_\nu-\tilde L_\nu)
\left[4fP^{\mu\nu }{}_\lambda+\kappa\Delta_\lambda{}^{\mu\nu}\right]=0\,,
\end{align}
where 
\[\Delta_\lambda{}^{\mu\nu}=-\frac2{\sqrt{-g}}\frac{\delta(\sqrt{-g}\mathcal L_M)}{\delta\Gamma^\lambda{}_{\mu\nu}}\,,\]is the hypermomentum tensor \cite{hyper}. 

The effective stress energy tensor $T^{\text{eff}}_{\mu\nu}$ is constructed using the relation
\begin{align*}
    f\m G_{\mu\nu}=\kappa T^{\text{eff}}_{\mu\nu}\,,
\end{align*}
where
\begin{equation} \label{T^eff}
 T^{\text{eff}}_{\mu\nu} 
 =  T_{\mu\nu}+ \frac 1{\kappa}\left[
        -2f'P^\lambda{}_{\mu\nu}\nabla_\lambda \phi
        +h\nabla_\mu\phi\nabla_\nu\phi 
        -\frac12hg_{\mu\nu}\nabla^\alpha\phi\nabla_\alpha\phi
        -\frac12 Ug_{\mu\nu}\right]\,.
\end{equation}
The additional part in (\ref{T^eff}) describes a source of fictitious dark energy that can drive the late-time acceleration driven by a negative pressure 
\begin{align}
T^{\text{DE}}_{\mu\nu}= \frac 1{f}\left[
        -2f'P^\lambda{}_{\mu\nu}\nabla_\lambda \phi
        +h\nabla_\mu\phi\nabla_\nu\phi 
        -\frac12hg_{\mu\nu}\nabla^\alpha\phi\nabla_\alpha\phi
        -\frac12 Ug_{\mu\nu}\right]\,.
\end{align}

In the present paper, we consider a perfect fluid type stress energy tensor given by
\begin{align}
T_{\mu\nu}=pg_{\mu\nu}+(p+\rho)u_\mu u_\nu
\end{align}
where $\rho$, $p$ and $u^\mu$ denote 
the energy density, pressure, and four velocity of the fluid, respectively.

\section{The cosmological fundamentals of non-metricity scalar-tensor theory}\label{sec1}
Following the cosmological principle,  the universe can be characterized by the FLRW spacetime,
which is homogeneous and isotropic on a large scale. 
The line element is given by 
\begin{align}\label{ds:RW}
ds^2=-dt^2+a^2\left(\frac{dr^2}{1-kr^2}+r^2d\theta^2+r^2\sin^2\theta d\phi^2\right)
\end{align}
where $a(t)$ is the scale factor of the universe; $H=\dot a/a$ 
is the Hubble parameter and the spatial curvature $k=0,+1,-1$ respectively 
modeled the universe of spatially flat, closed and open type.
Here the $\dot{(~)}$ denotes the derivative with respect to $t$.

There are three classes of affine connections that are compatible with the symmetric teleparallel framework, which are given as follows \cite{ FLRW/connection}:
\begin{align} \label{eqn:conn}
\Gamma^t{}_{tt}=&C_1, 
	\quad 					\Gamma^t{}_{rr}=\frac{C_2}{1-kr^2}, 
	\quad 					\Gamma^t{}_{\theta\theta}=C_2r^2, 
	\quad						\Gamma^t{}_{\phi\phi}=C_2r^2\sin^2\theta,								\notag\\
\Gamma^r{}_{tr}=&C_3, 
	\quad  	\Gamma^r{}_{rr}=\frac{kr}{1-kr^2}, 
	\quad		\Gamma^r{}_{\theta\theta}=-(1-kr^2)r, 
	\quad		\Gamma^r{}_{\phi\phi}=-(1-kr^2)r\sin^2\theta,												\notag\\
\Gamma^\theta{}_{t\theta}=&C_3, 
	\quad		\Gamma^\theta{}_{r\theta}=\frac1r,
	\quad		\Gamma^\theta{}_{\phi\phi}=-\cos\theta\sin\theta,										\notag\\
\Gamma^\phi{}_{t\phi}=&C_3, 
	\quad 	\Gamma^\phi{}_{r\phi}=\frac1r, 
	\quad 	\Gamma^\phi{}_{\theta\phi}=\cot\theta,
\end{align}
where $C_1$, $C_2$ and $C_3$ are temporal functions. In particular, when the corresponding curvature tensor is vanishing, the functions $C_1$, $C_2$ and $C_3$ are given by
\begin{enumerate}\label{casesConn}
\item[(I)] $C_1=\gamma$, $C_2=C_3=0$ and $k=0$, where $\gamma$ is a temporal function; or
\item[(II)] $C_1=\gamma+\dfrac{\dot\gamma}\gamma$, $C_2=0$, $C_3=\gamma$ and $k=0$,
             where $\gamma$ is a nonvanishing temporal function; or 
\item[(III)] $C_1=-\dfrac k\gamma-\dfrac{\dot\gamma}{\gamma}$, $C_2=\gamma$, $C_3=-\dfrac k\gamma$ and $k=0,\pm1$,
             where $\gamma$ is a nonvanishing temporal function.
\end{enumerate} 
As central to the present investigation, we focus on the connection class III which enables us to formulate the equations of motion in both spatially flat as well as spatially curved (open and closed type) FLRW spacetimes\footnote{Connection class I leads to a gravitational theory equivalent to scalar-torsion theory and the detailed cosmological implication emerged from the Connection class II has been studied in \cite{murtazascalar}.}. The corresponding Friedmann type equations of pressure and energy density are given as,

\begin{align}\label{peq}
\kappa p
=\; & f\left(-2\dot H-3H^2-\frac k{a^2}\right)
    +\frac12\dot f\left(-3\frac {k}{\gamma} +\frac{\gamma}{a^2}-4H\right)
    -\frac12h\dot \phi^2    +\frac12U\,,\\
    \kappa\rho
=\; & f\left(3H^2+3\dfrac k{a^2}\right)
    +\frac12\dot f\left(-3\frac {k}{\gamma} -3\frac{\gamma}{a^2}\right)
     -\frac12h\dot \phi^2    -\frac12U\,.    
    \end{align}

The scalar field equation (\ref{eqn:FE2}) yields
\begin{align}\label{fieldeq}
\left(-6H^2+3\frac {k}{\gamma}\left\{\frac{\dot{\gamma}}{\gamma}-3H\right\}
+3\left\{\frac{\dot{\gamma}}{a^2}+\frac{H\gamma}{a^2}+\frac{2k}{a^2}\right\}\right)f'
-h'\dot \phi^2 -2h(\ddot\phi+3H\dot \phi)-U'=0.
\end{align}
The equations of motion for the connection field equation (\ref{eqn:FE3}) reads 
\begin{align} \label{conneq}
0
=-\frac32\left[
    \dot f\left(3\frac{k}{\gamma} H+2\frac{\dot{\gamma}}{a^2}+\frac{H\gamma}{a^2}\right)
    +\ddot f\left(\frac{k}{ \gamma}+\frac{\gamma}{a^2}\right)\right].
\end{align}
and $\gamma$ is not a gauge parameter anymore.

An interesting feature of these cosmological models is that they allow for a minisuperspace formulation to simplify the study of cosmological dynamics by reducing the full degrees of freedom to a small set of variables. Specifically, for each connection, one can construct a Lagrangian function whose variation reproduces the corresponding field equations and exact solutions.
For the third connection $\Gamma_{III}$, the Lagrangian function can be given as
\begin{equation}
\mathcal{L}({\Gamma_{III}}) = f(\phi)\left(-3a 
\dot{a}^2 - \frac{3}{2} \frac{a \dot{\phi}}{\dot{\Psi}}+\frac{h(\phi)\dot\phi^2a^3}{2}\right) -\frac{U(\phi)a^3}{2}, ~~where ~\dot\Psi=\frac{1}{\gamma}
\end{equation}

\section{The formulation of the dynamical system and cosmological implications}\label{sec2}
We define the dimensionless dependent variables in the context of H-normalization \cite{Wands1998}
\begin{align}
\label{variables}
    x &= \frac{\dot{\phi}}{\sqrt{6f}H},
    y = \frac{U}{3H^2f}, 
    \Omega^k = \frac{k}{a^2H^2}, 
    z = \frac{\gamma}{a^2H},
    \Omega = \frac{\kappa\rho}{3H^2f}, \notag \\
    \lambda &= \frac{U' \sqrt{f}}{U}, 
    \mu = \frac{f'}{\sqrt{f}}, 
   \Delta  = \frac{U'' U}{U'^2}, 
   \Gamma = \frac{f''f}{f'^2}.
\end{align}

The constraint equation can be written as
\begin{align}
    \Omega=1+\Omega^k-\frac{\sqrt{3}}{\sqrt{2}}x \mu z-\frac{\sqrt{\frac{3}{2}}x \mu \Omega^k}{z}-h_0x^2-\frac{y}{2}.
\end{align}
From eq (\ref{peq}) by considering $p=0$, we can get following
\begin{align} {\label{dH}}
\frac{\dot{H}}{H^2}=
- \frac{3}{2} - \frac{3}{2} h_0 x^2 + \frac{3 y}{4} - \sqrt{6} \, x \, \mu + \frac{1}{2}\sqrt{\frac{3}{2}}x z \mu - \frac{\Omega^k}{2} - \frac{3 \sqrt{\frac{3}{2}} \, x \, \mu \, \Omega^k}{2 z} .
\end{align}
Using the equation (\ref{peq}), (\ref{fieldeq}) and (\ref{conneq}), the dynamical system equations can be given as,

\begin{align}
\bar{x} =\; & 
x \Bigg( 
24 \sqrt{6} h_0^2 x^3 z^4 
+ 6 h_0 x^2 z \mu \big(
  -((-8 + z) z^3) + 18 z^2 \Omega^k + 3 (\Omega^k)^{2}
\big) \nonumber \\
& + 3 z \big(
  -z^3 \big(8 \lambda y + (16 + (-10 + 3 y) z) \mu \big)
  + 2 z^2 (-2 - 3 y + z (-4 + z)) \mu \Omega^k \nonumber \\
& \quad + (-6 - 3 y + 4 z^2) \mu (\Omega^k)^2 
  + 2 \mu (\Omega^k)^3
\big) \nonumber + \sqrt{6} x \big(
  3 z^4 \big(-4 h_0 (2 + y) - z (-2 + z + 4 \Gamma ) \mu^2 \big) \\
&+ z^3 \big(8 h_0 z + 3 (4 + z - 8 \Gamma ) \mu^2 \big) \Omega^k  + 3 z (2 + 5 z - 4 \Gamma ) \mu^2 (\Omega^k)^2 
  + 9 \mu^2 (\Omega^k)^3
\big) 
\Bigg) \Bigg/ 4 z \left(4 \sqrt{6} h_0 x z^3 + 3 \mu (z^2 + \Omega^k)^2 \right),\\
\bar{z} =\; &
24 \sqrt{6} h_0^2 x^3 z^4 
- 6 h_0 x^2 z \mu \Big(
    z^3 (z + 8 (-2 + \Gamma )) 
    + 2 z (-9 z + 4 \Gamma ) \Omega^k 
    - 3 (\Omega^k)^2
\Big) \nonumber \\
& + 3 z (z^2 + \Omega^k) \Big(
    z \big(4 \lambda y + (8 - 3(2 + y) z) \mu\big)
    + (-2 - 3 y + 2 z (2 + z)) \mu \Omega^k 
    + 2 \mu (\Omega^k)^2
\Big) \nonumber \\
& + \sqrt{6} x \Big(
    z^4 \big(8 h_0 - 12 h_0 y - 3 (-4 + z) z \mu^2 \big) 
    + z^3 \big(8 h_0 z + 3 (8 + z) \mu^2 \big) \Omega^k \nonumber \\
& \qquad + 3 z (4 + 5 z) \mu^2 (\Omega^k)^2 
    + 9 \mu^2 (\Omega^k)^3
\Big)\Bigg/ \left(16 \sqrt{6} h_0 x z^3 + 12 \mu (z^2 + \Omega^k)^2\right),
\end{align}
\begin{align}
\bar{y}=\frac{
y \left(
6 h_0 x^2 z + 
z (6 - 3 y + 2 \Omega^k) + 
\sqrt{6} x 
  (-z^2 \mu + 
  2 z (\lambda + \mu) + 
  3 \mu \Omega^k)
\right)
}{
2 z
},
\end{align}

\begin{align}
\bar{\Omega^k}=
\frac{
\Omega^k \left(
6 h_0 x^2 z + 
z (2 - 3 y + 2 \Omega^k) + 
\sqrt{6} x \mu (
  -(-4 + z) z + 
  3 \Omega^k)
\right)
}{
2 z
},
\end{align}

\begin{align}{\label{lambda:new}}
 \overline{\lambda}=\sqrt{\frac{3}{2}}\, \lambda x \left(
2 \lambda (-1 + \Delta) + \mu
\right),
\end{align}
\begin{align}\label{zeta:eq}
 \overline{\mu}=\sqrt{\frac{3}{2}}\, x \left(-1 + 2 \Gamma  \right) \mu^2.
\end{align}
Here $\overline{(.)}$ represents derivative with respect to $N=lna$ or $d/dN$. Also, we consider a pressurless dust era in this study.

 The  deceleration parameter $q$ and the equation of state parameter for total fluid $w_{eff}$ can be expressed in terms of variables as
\begin{align}
q=\frac{
    z \left(2 + 6 h_0 x^2 - 3 y + 2 \Omega^k \right) + 
    \sqrt{6} \, x \mu \left(-((-4 + z) z) + 3 \Omega^k \right)
}{
    4z
},
\end{align}
\begin{align}
w_{eff}=h_0 x^2 + \frac{1}{6} \left(
    -3 y + 2 \Omega^k + 
    \frac{
        \sqrt{6} \, x \mu \left( -((-4 + z) z) + 3 \Omega^k \right)
    }{z}
\right).
\end{align}
Since the only functions for scalar fields are $\mu$ and $\lambda$. This yields four distinct scenarios for analysis. We write the dynamical equations for each of the case, but specialise ($h_0=1$) for the fixed point analysis.

\subsection{$\lambda=\lambda_0$, $\mu=\mu_0$}\label{case I}
Let us start with the case where both $\lambda$ and $\mu$ are constant. From (\ref{variables}), this leads to a coupling function $f(\phi)=\frac{\mu_0^2 \phi^2}{4}$ and a potential $U(\phi)=U_{0} \phi^{\frac{2 \lambda_0}{\mu_0}}$, with $U_{0}$ being a constant. Under this setting, our dynamical system reduces to four dimensions, and we have the following autonomous system,
\begin{align} \label{case1_eqx}
\bar{x} =\; & 
x \Bigg( 
24 \sqrt{6} h_0^2 x^3 z^4 
+ 6 h_0 x^2 z \mu_0 \big(
  -((-8 + z) z^3) + 18 z^2 \Omega^k + 3 (\Omega^k)^{2}
\big) \nonumber \\
& + 3 z \big(
  -z^3 \big(8 \lambda_0 y + (16 + (-10 + 3 y) z) \mu_0 \big)
  + 2 z^2 (-2 - 3 y + z (-4 + z)) \mu_0 \Omega^k \nonumber \\
& \quad + (-6 - 3 y + 4 z^2) \mu_0 (\Omega^k)^2 
  + 2 \mu_0 (\Omega^k)^3
\big) \nonumber + \sqrt{6} x \big(
  3 z^4 \big(-4 h_0 (2 + y) - z^2 \mu_0^2 \big) \\
&+ z^3 \big(8 h_0 z + 3 z \mu_0^2 \big) \Omega^k  + 15 z^2\mu_0^2 (\Omega^k)^2 
  + 9 \mu_0^2 (\Omega^k)^3
\big) 
\Bigg) \Bigg/ 4 z \left(4 \sqrt{6} h_0 x z^3 + 3 \mu_0 (z^2 + \Omega^k)^2 \right),\\
\bar{z} =\; & 
24 \sqrt{6} h_0^2 x^3 z^4 
- 6 h_0 x^2 z \mu_{0} 
\left( 
(-12 + z) z^3 
+ 2(2 - 9z) z \Omega^k 
- 3 (\Omega^k)^2 
\right) \nonumber \\
& 
+ 3 z (z^2 + \Omega^k) \left(
z \left(4 \lambda_0 y + (8 - 3(2 + y) z) \mu_{0} \right)
+ (-2 - 3y + 2z(2 + z)) \mu_{0} \Omega^k 
+ 2 \mu_{0} (\Omega^k)^2
\right) \nonumber \\
& 
+  \sqrt{6} x \Big(
z^4 \left(8 h_0 - 12 h_0 y - 3(-4 + z) z \mu_{0}^2 \right)
+ z^3 \left(8 h_0 z + 3(8 + z) \mu_{0}^2 \right) \Omega^k \nonumber \\
& 
+ 3 z (4 + 5z) \mu_{0}^2 (\Omega^k)^2
+ 9 \mu_{0}^2( \Omega^k)^3
\Big) \Big/ \left(16 \sqrt{6} h_0 x z^3 + 12 \mu_{0} (z^2 + \Omega^k)^2\right),
\end{align}

\begin{equation}
\bar{y}=\frac{
y \left(
6 h_0 x^2 z + 
z (6 - 3 y + 2 \Omega^k) + 
\sqrt{6} x 
  (-z^2 \mu_0 + 
  2 z (\lambda_0 + \mu_0) + 
  3 \mu_0 \Omega^k)
\right)
}{
2 z
},
\end{equation}

\begin{equation} \label{case1_eqOmega}
\bar{\Omega^k}=
\frac{
\Omega^k \left(
6 h_0 x^2 z + 
z (2 - 3 y + 2 \Omega^k) + 
\sqrt{6} x \mu_0 (
  -(-4 + z) z + 
  3 \Omega^k)
\right)
}{
2 z
}.
\end{equation}
The stability analysis of the critical points (CPs), namely $P_3$, $P_5$, $P_6$, and $P_8$, is not straightforward due to the strong non-linearity of the system. To proceed, we restrict our analysis to specific and physically meaningful values of the parameters $\mu_0$ and $\lambda_0$. In particular, we consider the following two limiting cases:
\begin{itemize}
\item When $\lambda \to 0$, in this limit, the scalar potential becomes effectively constant, thereby mimicking the behavior of a cosmological constant.
\item When $\mu \to 0$, this implies that the coupling function becomes constant, reducing the model to a minimally coupled scalar field theory. Since our focus is on non-minimally coupling scenarios, we avoid the vanishing value of $\mu$.
\end{itemize}
Based on these considerations, we carry out the stability analysis for the specific values $\mu_0 =\set {-1,1}$ and $\lambda_0=0$ for the aforementioned critical points.

\begin{table}[h]
    \centering
    \resizebox{\textwidth}{!}{ 
    \begin{tabular}{|c|c|c|c|c|}
        \hline
        Critical point  &$(x,y,z,\Omega^k)$ &Existence &$w_{eff}$ & $q$ \\ \hline
       $P_1$ & $\left( 0,0,\frac{4}{3}, 0 \right)$& Always & $0$&$\frac{1}{2}$ \\ \hline
         $P_2$ & $\left( 0,2,\frac{2(\lambda_0+\mu_0)}{3\mu_0}, 0 \right)$ & $\mu_0\neq 0$&$-1$&$-1$\\ \hline
          $P_3$ & $\left(\frac{5\sqrt{\frac{2}{3}}}{\mu_0-3\lambda_0},\frac{4(50-9\lambda_0^2-24\lambda_0 \mu_0+9\mu_0^2)}{3(3\lambda_0-\mu_0)^2},\frac{10-3\lambda_0^2-2\lambda_0\mu_0+\mu_0^2}{(3\lambda_0-\mu_0)\mu_0},0 \right)$ & $\mu_0 \neq 0 \land \mu_0 \in \mathbb{R} \land \lambda_0<\frac{\mu_0}{3}$&$-\frac{100-33\lambda_0^2+2\lambda_0\mu_0+3\mu_0^2}{3(-3\lambda_0+\mu_0)^2}$&$\frac{-50+(3\lambda_0-\mu_0)(7\lambda_0+\mu_0)}{(-3\lambda_0+\mu_0)^2}$\\ \hline
           $P_4$ & $\left(\frac{1}{\sqrt{6}\mu_0},0,\frac{1+2\mu_0^2}{\mu_0^2},0\right)$ & $\mu_0>0$&$\frac{1}{3}$&$1$\\ \hline
            $P_5$ & $\left(-\frac{\mu_0^2+\sqrt{\mu_0^2(2+\mu_0^2)}}{\sqrt{6}\mu_0},0,\frac{2(2\mu_0^2-\sqrt{2\mu_0^2+\mu_0^4})}{3\mu_0^2},0 \right)$ & $\mu_0<0$&$\frac{1}{9}(1-2\mu_0^2-2\sqrt{\mu_0^2(2+\mu_0^2)})$&$\frac{1}{3}(2-\mu_0^2-\sqrt{\mu_0^2(2+\mu_0^2)})$\\ \hline
             $P_6$ & $\left(\frac{-\mu_0^2+\sqrt{\mu_0^2(2+\mu_0^2)}}{\sqrt{6}\mu_0},0,\frac{2(2\mu_0^2+\sqrt{2\mu_0^2+\mu_0^4})}{3\mu_0^2},0 \right)$ & $\mu_0>0$&$\frac{1}{9}(1-2\mu_0^2+2\sqrt{\mu_0^2(2+\mu_0^2)})$&$\frac{1}{3}(2-\mu_0^2+\sqrt{\mu_0^2(2+\mu_0^2)})$\\ \hline
               $P_7$ & $\left( 0,0,-\frac{1}{2}, -1 \right)$ &Always&$-\frac{1}{3}$ & $0$\\ \hline
             $P_{8}$ & $\left(x,0,\frac{-120-8x^2-197\sqrt{6}x \mu_0+\sqrt{6}x^3\mu_0-492x^2\mu_0^2-48\sqrt{6}x^3\mu_0^3 }{120},\frac{-240-696\sqrt{6}x\mu_0+x^2(336-6807\mu_0^2)-123x^4\mu_0^2(-1+48\mu_0^2)-2\sqrt{6}x^3\mu_0(148+1875\mu_0^2)}{240} \right)$ &$ x$ arbitrary&$w_{eff}{P_8}$&$q_{}{P_8}$\\ \hline
    \end{tabular}
    }
    \caption{Critical points and their physical properties.}
    \label{case1table1}
\end{table}

\begin{table}[h]
\centering
\begin{tabular}{|c|c|l|}
\hline
Critical point & Eigenvalues & Stability \\ \hline
$P_1$ & $\left( -\frac{3}{2},-\frac{1}{2},1,3 \right)$ & saddle \\ \hline
$P_2$ & $\left(-5,-3,-3,-2\right)$ & stable \\ \hline
$P_{3_{(\lambda_0=0, \mu_0=1)}}$ & $\left(\frac{2}{7}(-14-\sqrt{22911}),\frac{2}{7}(-14+\sqrt{22911}),8,-3 \right)$ & saddle \\ \hline
$P_4$ & $\left(2,\frac{\lambda_0+3\mu_0}{\mu_0},A^{+},A^{-}\right)$ & saddle \\ \hline
$P_{5_{(\lambda_0=0, \mu_0=-1)}}$ & $\left(\frac{1}{3}(11+\sqrt{3}),-\frac{1}{3(-2+\sqrt{3})},-\frac{2}{3}(-1+\sqrt{3}),\frac{2(5-3\sqrt{3})}{3(-2+\sqrt{3})}\right)$ & saddle \\ \hline
$P_{6_{(\lambda_0=0, \mu_0=1)}}$ & $\left(\frac{1}{3}(11-\sqrt{3}),\frac{2(-5-3\sqrt{3})}{3(2+\sqrt{3})},\frac{2}{3}(1+\sqrt{3}),\frac{1}{3(2+\sqrt{3})}\right)$ & saddle \\ \hline
$P_7$ & $\left(-2,-1,2,2 \right)$ & saddle \\ \hline
$P_8$ & $\left(\lambda_1(x),\lambda_2(x),\lambda_3(x),\lambda_4(x)\right)$ & \textbf{Case I:} $\mu_0=1,\lambda_0=0$ \\
 & & \quad -- unstable for $x \ge 0.75$ \\
 & & \quad -- stable for $x \le 0.74$ \\
 & & \textbf{Case II:} $\mu_0 = -1,\lambda_0=0$ \\
 & & \quad -- unstable for $x < 0$ \\
 & & \quad -- stable for $x \ge 0$ \\ \hline
\end{tabular}
 \caption{Stability analysis of the critical points. $A^{\pm}=\frac{-7\mu_0^3 - 20\mu_0^5 - 12\mu_0^7 \pm \sqrt{-7\mu_0^4 + \mu_0^6 + 160\mu_0^8 + 44\mu_0^{10} + 432\mu_0^{12} + 144\mu_0^{14}}}
         {\mu_0^3(1 + 2\mu_0^2)(7 + 6\mu_0^2)}$.}
\label{case1table2}
\end{table}

\begin{figure}[htbp]
    \centering
    
    \begin{subfigure}[t]{0.31\linewidth}
        \centering
        \includegraphics[width=\linewidth]{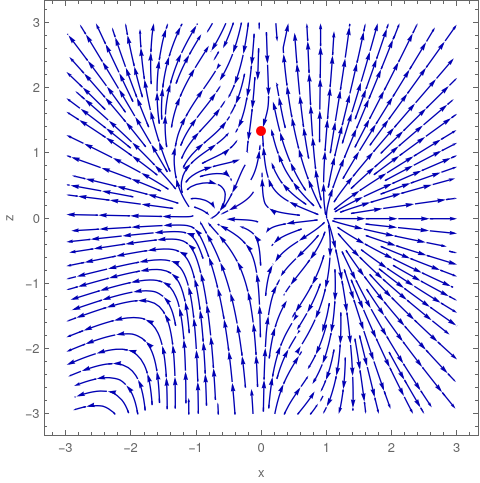}
        \caption{Phase-space portrait of the attractor point $P_2$ for $\lambda_0=1$ and $\mu_0=1$. }
        \label{fig:sub1}
    \end{subfigure}
    \hspace{0.01\linewidth}
    \begin{subfigure}[t]{0.31\linewidth}
        \centering
        \includegraphics[width=\linewidth]{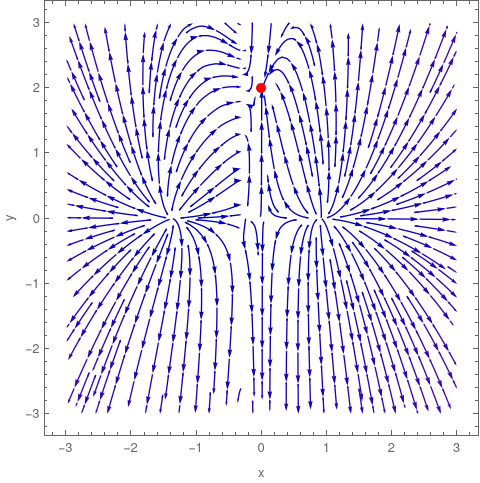}
        \caption{Phase-space portrait of the attractor point $P_2$ for $\lambda_0=0$, $\mu_0=1$ and $z=1$. }
        \label{fig:sub2}
    \end{subfigure}
    \hspace{0.01\linewidth}
    \begin{subfigure}[t]{0.31\linewidth}
        \centering
        \includegraphics[width=\linewidth]{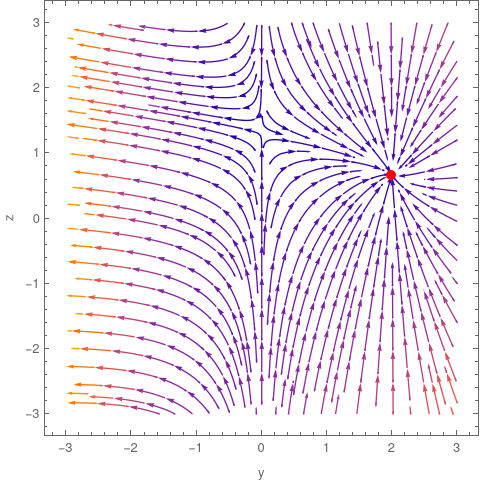}
        \caption{Phase-space portrait of the attractor point $P_2$ for $\lambda_0=0$ and $\mu_0=1$. }
        \label{fig:sub3}
    \end{subfigure}
    
    \vspace{0.6cm} 
    \begin{subfigure}[t]{0.48\linewidth}
        \centering
        \includegraphics[width=\linewidth]{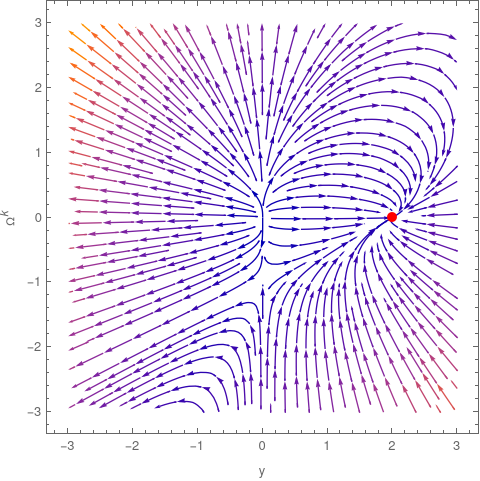}
        \caption{Phase-space portrait of the attractor point $P_2$ for $\lambda_0=0$, $\mu_0=-1$ and $z=1$.}
        \label{fig:sub4}
    \end{subfigure}
    \hspace{0.03\linewidth}
    \begin{subfigure}[t]{0.48\linewidth}
        \centering
        \includegraphics[width=\linewidth]{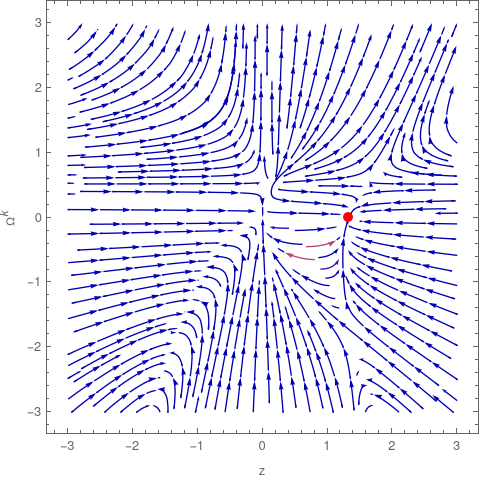}
        \caption{Phase-space portrait of the attractor point $P_2$ for $\lambda_0=1$ and $\mu_0=1$.}
        \label{fig:sub5}
    \end{subfigure}
    
    \caption{Phase-space portraits of attractor solution for (\textbf{Case \ref{case I}}).}
    \label{case1_fig1}
\end{figure}

\begin{figure}[htbp]
    \centering
    
    \begin{subfigure}[t]{0.48\textwidth}
        \centering
        \includegraphics[width=\linewidth]{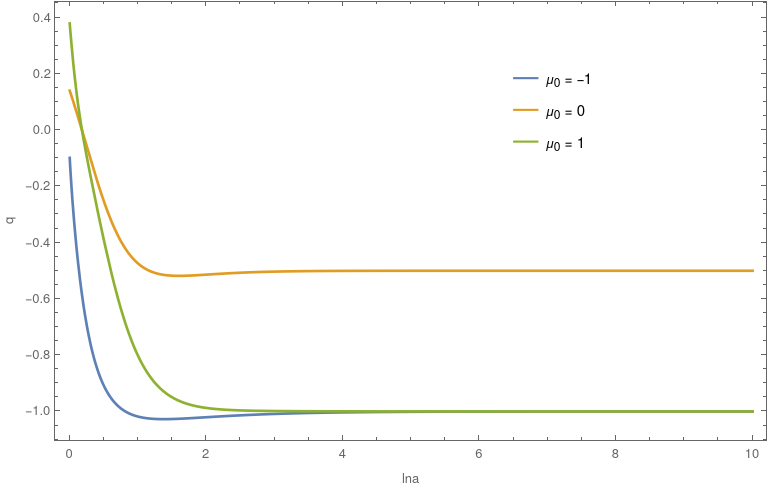}
        \caption{Deceleration parameter for $\mu_0$ values.}
        \label{fig:sub1}
    \end{subfigure}
    \hfill
    \begin{subfigure}[t]{0.48\textwidth}
        \centering
        \includegraphics[width=\linewidth]{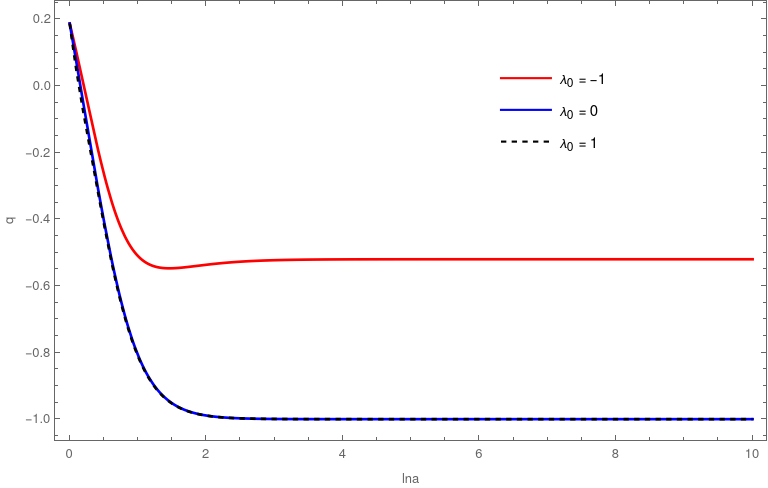}
        \caption{Deceleration parameter for $\lambda_0$ values.}
        \label{fig:sub2}
    \end{subfigure}
    
    \caption{Qualitative evolution of the deceleration parameter of the dynamical system (\ref{case1_eqx})-(\ref{case1_eqOmega})
for different values of $\lambda_0$ and $\mu_0$, with initial conditions ($x[0]=0.1,~y[0]=0.5 ,~z[0]=0.1,~\Omega^{k}[0]=0$) for
(\textbf{Case \ref{case I}}).}
    \label{case1_fig2}
\end{figure}

A detailed analysis of each critical point is presented in Tables \ref{case1table1} and \ref{case1table2}. The stationary points $P_1$ to $P_6$ correspond to a spatially flat universe. Among them, the fixed point $P_1$  represents a matter-dominated epoch and exhibits saddle behavior, indicating its role as a transient phase in cosmic evolution. The critical point $P_2$  is the only stable attractor in the phase space and corresponds to a future de Sitter universe, consistent with late-time cosmic acceleration. The fixed points $P_3, P_5$, and $P_6$ are saddle points with parameter-dependent physical characteristics, while $P_4$ describes a decelerating universe and also exhibits saddle-like dynamics.  The critical point $P_7$ and $P_8$ correspond to a non-flat universe. The fixed point $P_7$ is characterized by a vanishing deceleration parameter. As a result, this point is not physically viable if the scalar field is interpreted as a dark energy candidate. The CP $P_8$ describes decelerated universe for $\left(\mu_0 \leq -\sqrt{\frac{2}{3}} \land (\frac{1}{62}(13-3\sqrt{129}))<x<0\right)$ and accelerated universe when $\left((-\sqrt{\frac{2}{3}}<\mu_0<-\frac{1}{\sqrt{2}})  \land 0<x<(\frac{1}{62}(13-3\sqrt{129}))\right)$. The 2D phase space portraits for the attractor solution using different values of the parameters $\lambda_0$ and $\mu_0$ against different variables are displayed in Fig \ref{case1_fig1}. The qualitative evolution of the deceleration parameter $q$ for the autonomous system (\ref{case1_eqx})-(\ref{case1_eqOmega}), for a different set of values of free parameters, is also depicted in Fig \ref{case1_fig2}. A similar analysis, considering the second class of non-coincident gauge under the same framework in a spatially flat FLRW spacetime, has been carried out in \cite{murtazascalar}. Therefore, it is worthwhile to briefly compare our curvature free critical points with those discussed therein.\footnote{In \cite{murtazascalar}, three fixed points were obtained, whereas in our case, we find six stationary points corresponding to the spatially flat universe. Among the critical points in \cite{murtazascalar}, one described a matter-dominated universe and another yielded a de Sitter solution, both consistent with our critical points $P_1$ and $P_2$. The third fixed point reported was parameter-dependent. In contrast, our analysis reveals three parameter-dependent critical points and one describing a decelerating universe, thereby making our results richer.}

\scalebox{0.8}{
\begin{minipage}{\linewidth}
\begin{align*}
w_{eff}{P_8}=
& -\Bigg(9600 + x \Bigg(5200 \sqrt{6} \mu_0 + x \Bigg( 
-41600 + 51624 \mu_0^2 + 39 \sqrt{6} x^5 \mu_0^3 (1 - 48 \mu_0^2)^2 \\
& \qquad + 3 \sqrt{6} x \mu_0 (-12496 + 36953 \mu_0^2) 
+ 24 x^4 \mu_0^2 (-37 + 705 \mu_0^2 + 51408 \mu_0^4) + 8 x^2 (-352 - 19518 \mu_0^2 + 136251 \mu_0^4) \\
& \qquad
+ 2 \sqrt{6} x^3 \mu_0 (560 + 18423 \mu_0^2 + 319428 \mu_0^4)
\Bigg)\Bigg)\Bigg)\Bigg/240 \Bigg(120 + x \Bigg(197 \sqrt{6} \mu_0 + x \Bigg(
8 + 492 \mu_0^2 + \sqrt{6} x \mu_0 (-1 + 48 \mu_0^2)
\Bigg)\Bigg)\Bigg),
\end{align*}
\end{minipage}
}

\scalebox{0.8}{
\begin{minipage}{\linewidth}
\begin{align*}
q_{}{P_8}=& x \Bigg(10560 \sqrt{6} \mu_0 + x \Bigg(
42240 - 12264 \mu_0^2 
+ \sqrt{6} x \mu_0 (37408 - 107019 \mu_0^2) \\
& \qquad 
- 39 \sqrt{6} x^5 \mu_0^3 (1 - 48 \mu_0^2)^2 
+ 8 x^2 (352 + 19518 \mu_0^2 - 136251 \mu_0^4) 
- 24 x^4 \mu_0^2 (-37 + 705 \mu_0^2 + 51408 \mu_0^4) \\
& \qquad 
- 2 \sqrt{6} x^3 \mu_0 (560 + 18423 \mu_0^2 + 319428 \mu_0^4)
\Bigg)\Bigg)\Bigg/ 160 \Bigg(120 + x \Bigg(
197 \sqrt{6} \mu_0 + x \Big(
8 + 492 \mu_0^2 + \sqrt{6} x \mu_0 (-1 + 48 \mu_0^2)
\Big)
\Bigg)\Bigg).
\end{align*}
\end{minipage}
}

\subsection{$\lambda=\lambda_0$ and $\mu$ is variable}\label{case II}
The most natural choice for the coupling and potential functions is either an exponential form or a power law. In this case, we can choose either; however, considering an exponential potential leads to a constant coupling function, and thus we discard that option. Moreover, the power law potential has already been analyzed in the previous scenario \ref{case I}. Therefore, we assume an exponential coupling function, $f(\phi)=f_0 e^{\alpha \phi}$ and from (\ref{variables}) we can obtain $U(\phi)=exp (\frac{-2\lambda_0}{\sqrt{f_0}\alpha}e^{-\frac{\alpha \phi}{2}})$. So, the autonomous dynamical system for this case is provided as,
\begin{align}\label{case2_xeq}
\bar{x} =\; & 
x \Bigg( 
24 \sqrt{6} h_0^2 x^3 z^4 
+ 6 h_0 x^2 z \mu \big(
  -((-8 + z) z^3) + 18 z^2 \Omega^k + 3 (\Omega^k)^{2}
\big) \nonumber \\
& + 3 z \big(
  -z^3 \big(8 \lambda_0 y + (16 + (-10 + 3 y) z) \mu \big)
  + 2 z^2 (-2 - 3 y + z (-4 + z)) \mu \Omega^k \nonumber \\
& \quad + (-6 - 3 y + 4 z^2) \mu (\Omega^k)^2 
  + 2 \mu (\Omega^k)^3
\big) \nonumber + \sqrt{6} x \big(
  3 z^4 \big(-4 h_0 (2 + y) - z (2 + z ) \mu^2 \big) \\
&+ z^3 \big(8 h_0 z + 3 ( z - 4 ) \mu^2 \big) \Omega^k   + 3 z ( 5 z-  2  ) \mu^2 (\Omega^k)^2 
  + 9 \mu^2 (\Omega^k)^3
\big) 
\Bigg) \Bigg/ 4 z \left(4 \sqrt{6} h_0 x z^3 + 3 \mu (z^2 + \Omega^k)^2 \right),\\
\bar{z} =\; &
24 \sqrt{6} h_0^2 x^3 z^4 
- 6 h_0 x^2 z \mu \Big(
    z^3 (z -8) 
    + 2 z (-9 z + 4  ) \Omega^k 
    - 3 (\Omega^k)^2
\Big) \nonumber \\
& + 3 z (z^2 + \Omega^k) \Big(
    z \big(4 \lambda_0 y + (8 - 3(2 + y) z) \mu\big)
    + (-2 - 3 y + 2 z (2 + z)) \mu \Omega^k 
    + 2 \mu (\Omega^k)^2
\Big) \nonumber \\
& + \sqrt{6} x \Big(
    z^4 \big(8 h_0 - 12 h_0 y - 3 (-4 + z) z \mu^2 \big) 
    + z^3 \big(8 h_0 z + 3 (8 + z) \mu^2 \big) \Omega^k \nonumber \\
& \qquad + 3 z (4 + 5 z) \mu^2 (\Omega^k)^2 
    + 9 \mu^2 (\Omega^k)^3
\Big)\Bigg/ \left(16 \sqrt{6} h_0 x z^3 + 12 \mu (z^2 + \Omega^k)^2\right),
\end{align}
\begin{align}
\bar{y}=\; &\frac{
y \left(
6 h_0 x^2 z + 
z (6 - 3 y + 2 \Omega^k) + 
\sqrt{6} x 
  (-z^2 \mu + 
  2 z (\lambda_0 + \mu) + 
  3 \mu \Omega^k)
\right)
}{
2 z
},\\
\bar{\Omega^k}=\; &
\frac{
\Omega^k \left(
6 h_0 x^2 z + 
z (2 - 3 y + 2 \Omega^k) + 
\sqrt{6} x \mu (
  -(-4 + z) z + 
  3 \Omega^k)
\right)
}{
2 z
},
\end{align}
\begin{align}\label{case2_mueq}
 \overline{\mu}=\; &\sqrt{\frac{3}{2}}\, x  \mu^2.
\end{align}

\begin{figure}[htbp]
    \centering
    
    \begin{subfigure}[t]{0.48\linewidth}
        \centering
        \includegraphics[width=\linewidth]{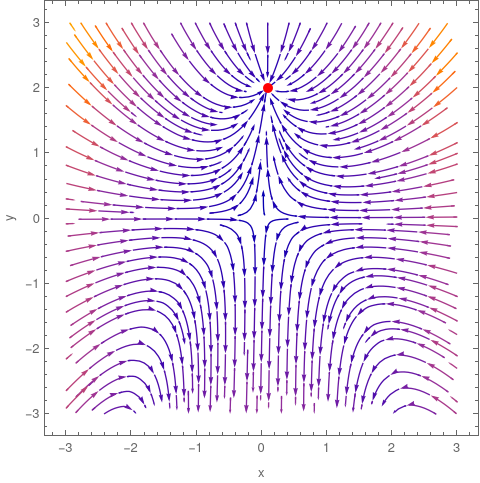}
        \caption{Phase-space portrait of the attractor point $P_2$ for $\lambda_0=1$ and $z=1$.}
        \label{fig:sub1}
    \end{subfigure}
    \hspace{0.03\linewidth}
    \begin{subfigure}[t]{0.48\linewidth}
        \centering
        \includegraphics[width=\linewidth]{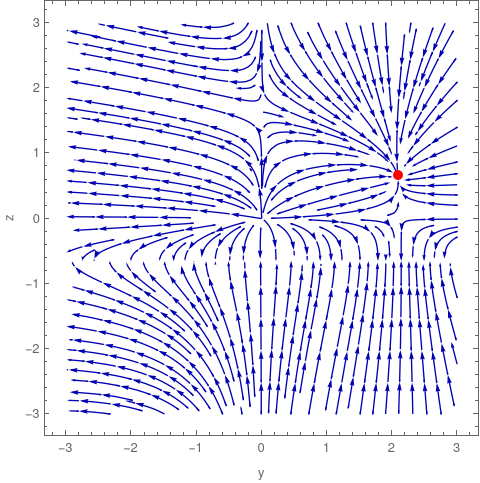}
        \caption{Phase-space portrait of the attractor point $P_2$ for $\lambda_0=-1$.}
        \label{fig:sub2}
    \end{subfigure}
    
    \vspace{0.6cm} 
    \begin{subfigure}[t]{0.48\linewidth}
        \centering
        \includegraphics[width=\linewidth]{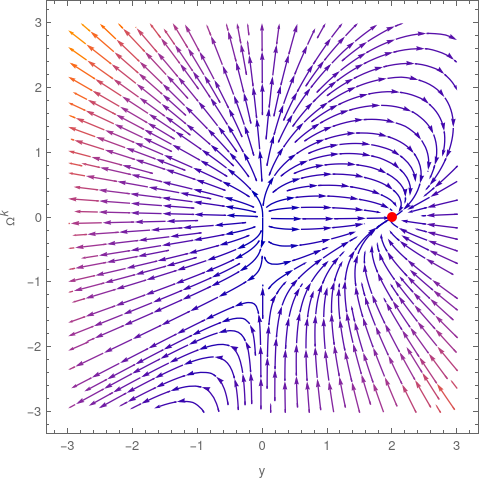}
        \caption{Phase-space portrait of the attractor point $P_2$ for $\lambda_0=1$ and $z=1$.}
        \label{fig:sub3}
    \end{subfigure}
    \hspace{0.03\linewidth}
    \begin{subfigure}[t]{0.48\linewidth}
        \centering
        \includegraphics[width=\linewidth]{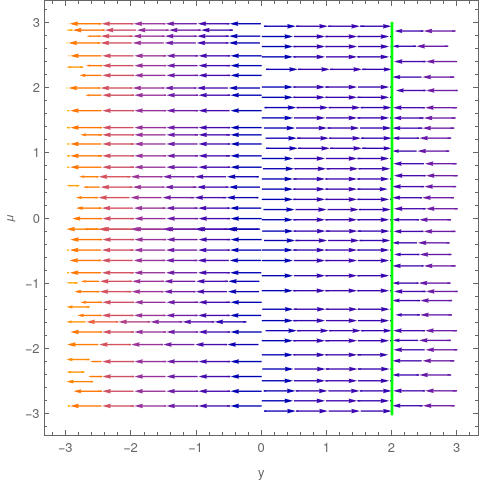}
        \caption{Phase-space portrait showing the vertical attractor line corresponding to $P_2$ for $\lambda_0=-1$ and $z=1$ .}
        \label{fig:sub4}
    \end{subfigure}
    
    \caption{Phase-space portraits of attractor solution for (\textbf{Case \ref{case II}}).}
    \label{case2_fig1}
\end{figure}

\begin{figure}[htbp]
    \centering
    \includegraphics[width=0.7\linewidth]{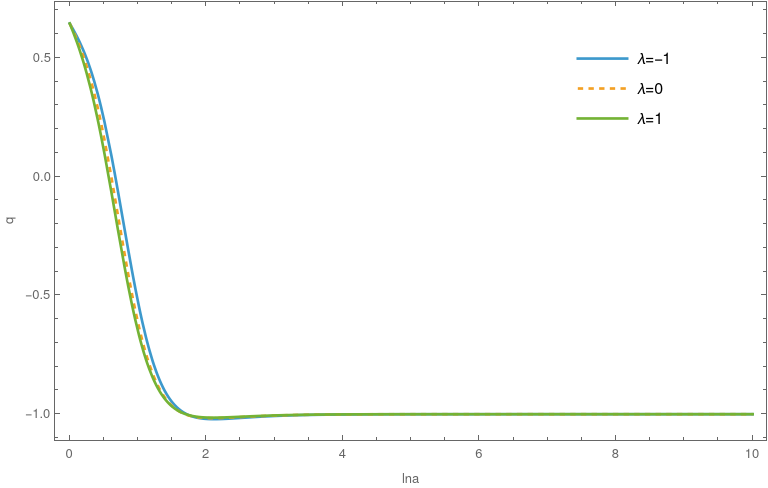}
    \caption{Qualitative evolution of the deceleration parameter of the dynamical system (\ref{case2_xeq})-(\ref{case2_mueq})
for different values of $\lambda_0$ with initial conditions ($x[0]=0.1,~y[0]=0.2 ,~z[0]=0.5,~\Omega^{k}[0]=0.3,~\mu[0]=0.4$) for
(\textbf{Case \ref{case II}}).}
    \label{case2_fig2}
\end{figure}

\begin{table}[h]
    \centering
    
    \begin{tabular}{|c|c|c|c|c|}
        \hline
        Critical point  &$(x,y,z,\Omega^k,\mu)$ &Existence &$w_{eff}$ & $q$\\ \hline
        
         $P_1$ & $\left( 0,0,\frac{4}{3}, 0 ,\mu\right)$&$\mu \neq 0$ & $0$&$\frac{1}{2}$ \\ \hline
         $P_2$ & $\left( 0,2,z, 0,\frac{2\lambda_0}{3z-2} \right)$ & $z\neq \frac{2}{3}$&$-1$&$-1$\\ \hline
           $P_3$ & $\left(-\frac{\lambda_0}{\sqrt{6}},\frac{8}{9},z,0,0 \right)$ & $\lambda_0=-\sqrt{\frac{10}{3}} \land z \neq 0$&$\frac{1}{9}$ &$\frac{2}{3}$ \\ \hline
           $P_4$ & $\left(-\sqrt{\frac{2}{3}}\lambda_0,\frac{8}{3},\pm i,1,0\right)$ & unphysical&- &- \\ \hline
           $P_5$ & $\left( 0,0,-\frac{1}{2}, -1, \mu \right)$ &$\mu \neq 0$&$-\frac{1}{3}$ & $0$\\ \hline

           \end{tabular}
    \caption{Critical points and their physical properties.}
    \label{case2table1}
\end{table}
\begin{table}[h]
    \centering
    \begin{tabular}{|c|c|c|}
        \hline
        Critical point  &Eigenvalues &Stability \\ \hline
       
         $P_1$ & $\left(0, -\frac{3}{2},-\frac{1}{2},1,3 \right)$& saddle \\ \hline
         $P_2$ & $\left(0,-5,-3,-3,-2\right)$ & stable\\ \hline
          $P_3$ & $\left(0,0,-\frac{4}{3},\frac{1}{3},\frac{4}{3}\right)$ & saddle\\ \hline
           $P_5$ & $\left(0,-2,-1,2,2 \right)$ &saddle\\ \hline
             \end{tabular}
    \caption{Stability analysis of the critical points.}
    \label{case2table2}
\end{table}
The description of each stationary point is provided in Tables \ref{case2table1} and \ref{case2table2}. We have three critical points $P_1$ to $P_3$ that correspond to a spatially flat universe.  Among the curvature-free spatially flat critical points, $P_1$ describes a matter-dominated universe with an unstable nature. The fixed point $P_2$ corresponds to a de Sitter solution. Although it possesses a zero eigenvalue, the application of the center manifold theorem \cite{Bahamonde2018b} shows that this point is stable, supporting the role as a viable late-time attractor. The fixed point $P_3$ represents a decelerating universe with a saddle nature. The critical points $P_4$ and $P_5$ correspond to a spatially non-flat universe, respectively. However, the CP $P_5$ is unphysical, and $P_5$ with $q=0$ makes it physically unacceptable as discussed earlier. In Fig \ref{case2_fig1}, the 2D phase space portraits of the attractor solution for various values of the parameter $\lambda_0$ with respect to different variables can be visualized. Similarly, Fig \ref{case2_fig2} illustrates the qualitative behavior of the deceleration parameter $q$ for different values of the free parameter $\lambda_0$. \footnote{ A related investigation in \cite{murtazascalar} identified four spatially flat equilibrium points, one corresponded to a matter-dominated era and another to a de Sitter phase, both in agreement with our fixed points $P_1$ and $P_2$. In addition, the analysis in \cite{murtazascalar} yielded a stationary point associated with a stiff-fluid solution and another depended on model parameters. By contrast, our third curvatureless fixed point gives a decelerating universe.}

\subsection{$\mu=\mu_0$ and $\lambda$ is variable}\label{case III}
In this particular case, from (\ref{variables}), the coupling function is given by $f(\phi)=\frac{\mu_0^2 \phi^2}{4}$. Here for the potential function, the most appropriate choice is an exponential form, $U(\phi)=U_0e^{\beta\phi}$, since the power law scenario has already been studied in \ref{case I}. The autonomous dynamical system corresponding to this case can be written as,
\begin{align}\label{case3_xeq}
\bar{x} =\; & 
x \Bigg( 
24 \sqrt{6} h_0^2 x^3 z^4 
+ 6 h_0 x^2 z \mu_0 \big(
  -((-8 + z) z^3) + 18 z^2 \Omega^k + 3 (\Omega^k)^{2}
\big) \nonumber \\
& + 3 z \big(
  -z^3 \big(8 \lambda y + (16 + (-10 + 3 y) z) \mu_0 \big)
  + 2 z^2 (-2 - 3 y + z (-4 + z)) \mu_0 \Omega^k \nonumber \\
& \quad + (-6 - 3 y + 4 z^2) \mu_0 (\Omega^k)^2 
  + 2 \mu_0 (\Omega^k)^3
\big) \nonumber + \sqrt{6} x \big(
  3 z^4 \big(-4 h_0 (2 + y) - z^2 \mu_0^2 \big) \\
&+ z^3 \big(8 h_0 z + 3z  \mu_0^2 \big) \Omega^k  + 15 z^2  \mu_0^2 (\Omega^k)^2 
  + 9 \mu_0^2 (\Omega^k)^3
\big) 
\Bigg) \Bigg/ 4 z \left(4 \sqrt{6} h_0 x z^3 + 3 \mu_0 (z^2 + \Omega^k)^2 \right),\\
\bar{z} =\; &
24 \sqrt{6} h_0^2 x^3 z^4 
- 6 h_0 x^2 z \mu_0 \Big(
    z^3 (z-12) 
    + 2 z (-9 z + 2 ) \Omega^k 
    - 3 (\Omega^k)^2
\Big) \nonumber \\
& + 3 z (z^2 + \Omega^k) \Big(
    z \big(4 \lambda y + (8 - 3(2 + y) z) \mu_0\big)
    + (-2 - 3 y + 2 z (2 + z)) \mu_0 \Omega^k 
    + 2 \mu_0 (\Omega^k)^2
\Big) \nonumber \\
& + \sqrt{6} x \Big(
    z^4 \big(8 h_0 - 12 h_0 y - 3 (-4 + z) z \mu_0^2 \big) 
    + z^3 \big(8 h_0 z + 3 (8 + z) \mu_0^2 \big) \Omega^k \nonumber \\
& \qquad + 3 z (4 + 5 z) \mu_0^2 (\Omega^k)^2 
    + 9 \mu_0^2 (\Omega^k)^3
\Big)\Bigg/ \left(16 \sqrt{6} h_0 x z^3 + 12 \mu_0 (z^2 + \Omega^k)^2\right),\\
\bar{y}=\; &
\frac{
y \left(
6 h_0 x^2 z + 
z (6 - 3 y + 2 \Omega^k) + 
\sqrt{6} x 
  (-z^2 \mu_0 + 
  2 z (\lambda + \mu_0) + 
  3 \mu_0 \Omega^k)
\right)
}{
2 z
},
\end{align}

\begin{align}
\bar{\Omega^k}=
\frac{
\Omega^k \left(
6 h_0 x^2 z + 
z (2 - 3 y + 2 \Omega^k) + 
\sqrt{6} x \mu_0 (
  -(-4 + z) z + 
  3 \Omega^k)
\right)
}{
2 z
},
\end{align}

\begin{align}{\label{case3_mueq}}
 \overline{\lambda}=\sqrt{\frac{3}{2}}\, \lambda x\mu_0.
\end{align}

\begin{table}[h]
    \centering
    \resizebox{\textwidth}{!}{ 
    \begin{tabular}{|c|c|c|c|c|}
        \hline
        Critical point  &$(x,y,z,\Omega^k,\lambda)$ &Existence &$w_{eff}$ & $q$ \\ \hline
       
         $P_1$ & $\left( 0,0,\frac{4}{3}, 0 ,\lambda\right)$& $\lambda$ arbitrary & $0$&$\frac{1}{2}$ \\ \hline
         $P_2$ & $\left( 0,2,z,0,\frac{(3z-2)\mu_0}{2} \right)$ & $z \neq 0$ &$-1$&$-1$\\ \hline
          $P_3$ & $\left(\frac{5\sqrt{\frac{2}{3}}}{\mu_0},\frac{4(50+9\mu_0^2)}{3\mu_0^2},\frac{-10-\mu_0^2}{\mu_0^2},0,0 \right)$ & $\mu_0 > 0$&$\frac{7}{3}$&$4$\\ \hline
           $P_4$ & $\left(\frac{1}{\sqrt{6}\mu_0},0,\frac{1+2\mu_0^2}{\mu_0^2},0,0\right)$ & $\mu_0>0$&$\frac{1}{3}$&$1$\\ \hline
            $P_5$ & $\left(-\frac{\mu_0^2+\sqrt{\mu_0^2(2+\mu_0^2)}}{\sqrt{6}\mu_0},0,\frac{2(2\mu_0^2-\sqrt{2\mu_0^2+\mu_0^4})}{3\mu_0^2},0,0 \right)$ & $\mu_0<0$&$\frac{1}{9}(1-2\mu_0^2-2\sqrt{\mu_0^2(2+\mu_0^2)})$&$\frac{1}{3}(2-\mu_0^2-\sqrt{\mu_0^2(2+\mu_0^2)})$\\ \hline
             $P_6$ & $\left(\frac{-\mu_0^2+\sqrt{\mu_0^2(2+\mu_0^2)}}{\sqrt{6}\mu_0},0,\frac{2(2\mu_0^2+\sqrt{2\mu_0^2+\mu_0^4})}{3\mu_0^2},0,0 \right)$ & $\mu_0>0$&$\frac{1}{9}(1-2\mu_0^2+2\sqrt{\mu_0^2(2+\mu_0^2)})$&$\frac{1}{3}(2-\mu_0^2+\sqrt{\mu_0^2(2+\mu_0^2)})$\\ \hline
              $P_7$ & $\left( 0,0,-\frac{1}{2}, -1,\lambda \right)$ &$\lambda$ arbitrary&$-\frac{1}{3}$ & $0$\\ \hline
              $P_8{\pm}$ & $\left(\pm \sqrt{\frac{2}{3}},\frac{8}{3},\pm i,1,\pm1 \right)$ &unphysical&-&-\\ \hline
              $P_9{\pm}$ & $\left(\frac{\sqrt{\frac{2}{3}}}{\mu_0},\frac{-4+9\mu_0^2\pm3\mu_0\sqrt{16+9\mu_0^2}}{3\mu_0^2}, \frac{-\mu_0^2\pm\mu_0\sqrt{16+9\mu_0^2}}{2\mu_0^2},\frac{-8-5\mu_0^2\pm\mu_0\sqrt{16+9\mu_0^2}}{2\mu_0^2},0\right)$ &$\mu_0>0$&$-\frac{1}{3}\pm\frac{\sqrt{16+9\mu_0^2}}{\mu_0}$&$\pm\frac{3\sqrt{16+9\mu_0^2}}{2\mu_0}$\\ \hline
               $P_{10}$ & $\left(x,0,\frac{-120-8x^2-197\sqrt{6}x \mu_0+\sqrt{6}x^3\mu_0-492x^2\mu_0^2-48\sqrt{6}x^3\mu_0^3 }{120},\frac{-240-696\sqrt{6}x\mu_0+x^2(336-6807\mu_0^2)-123x^4\mu_0^2(-1+48\mu_0^2)-2\sqrt{6}x^3\mu_0(148+1875\mu_0^2)}{240} ,0\right)$ &$ x$ arbitrary&$w_{eff}{P_{10}}$&$q_{}{P_{10}}$\\ \hline
    \end{tabular}
    }
    \caption{Critical points and their physical properties.}
    \label{case3table1}
\end{table}
\begin{table}[h]
\centering
\begin{tabular}{|c|c|l|}
\hline
Critical point & Eigenvalues & Stability \\ \hline
       
         $P_1$ & $\left( 0,-\frac{3}{2},-\frac{1}{2},1,3 \right)$& saddle \\ \hline
         $P_2$ & $\left(0,-5,-3,-3,-2\right)$ & stable\\ \hline
          $P_3$ & $\left(5,8,-\frac{3(-100\mu_0^3+20\mu_0^5+3\mu_0^7)}{\mu_0^3(10+\mu_0^2)(-10+3\mu_0^2)},B^{+},B^{-} \right)$ & saddle\\ \hline
           $P_4$ & $\left(\frac{1}{2},2,3,C^{+},C^{-}\right)$ & saddle\\ \hline
            $P_{5_{( \mu_0=-1)}}$ & $\left(\frac{1}{3}(11+\sqrt{3}),\frac{1}{2}(-1-\sqrt{3}),-\frac{1}{3(-2+\sqrt{3})},-\frac{2}{3}(-1+\sqrt{3}),\frac{2(5-3\sqrt{3})}{3(-2+\sqrt{3})}\right)$ & saddle\\ \hline
             $P_{6_{( \mu_0=1)}}$ &  $\left(\frac{1}{3}(11-\sqrt{3}),\frac{2(-5-3\sqrt{3})}{3(2+\sqrt{3})},\frac{2}{3}(1+\sqrt{3}),\frac{2}{3}(-1+\sqrt{3}),\frac{1}{3(2+\sqrt{3})}\right)$ & saddle\\ \hline
              $P_7$ & $\left(0,-2,-1,2,2 \right)$ &saddle\\ \hline
             $P_{9\pm_{( \mu_0=1)}}$ &  $\left(-33.97,-14.57,-13.45,-9,1\right)_+$
           \newline  $\left(28.23,15.39,2.68+4.47i,2.68-4.47i,1\right)_-$ & saddle\\ \hline
            $P_{10}$ & $\left(\lambda_1(x),\lambda_2(x),\lambda_3(x),\lambda_4(x),\lambda_5(x)\right)$ & \textbf{Case I:} $\mu_0=1$ \\
 & & \quad -- unstable for $x \ge 0.75$ \\
 & & \quad -- stable for $x \le 0.74$ \\
 & & \textbf{Case II:} $\mu_0 = -1$ \\
 & & \quad -- unstable for $x < 0$ \\
 & & \quad -- stable for $x \ge 0$ \\ \hline
   \end{tabular}
    
    \caption{Stability analysis of the critical points. $B^{\pm}=\frac{2(200\mu_0^3-40\mu_0^5-6\mu_0^7\pm\sqrt{2500000\mu_0^4+490000\mu_0^6-166000\mu_0^8-47800\mu_0^{10}-3870\mu_0^{12}-99\mu_0^{14})}}{\mu_0^3(10+\mu_0^2)(-10+3\mu_0^2)} ,C^{\pm}=\frac{-7\mu_0^3-20\mu_0^5-12\mu_0^7\pm\sqrt{-7\mu_0^4+\mu_0^6+160\mu_0^8+440\mu_0^{10}+432\mu_0^{12}+144\mu_0^{14}}}{\mu_0^3(1+2\mu_0^2)(7+6\mu_0^2)}$
    }
\label{case3table2}
\end{table}

\begin{figure}[htbp]
    \centering
    
    \begin{subfigure}[t]{0.31\linewidth}
        \centering
        \includegraphics[width=\linewidth]{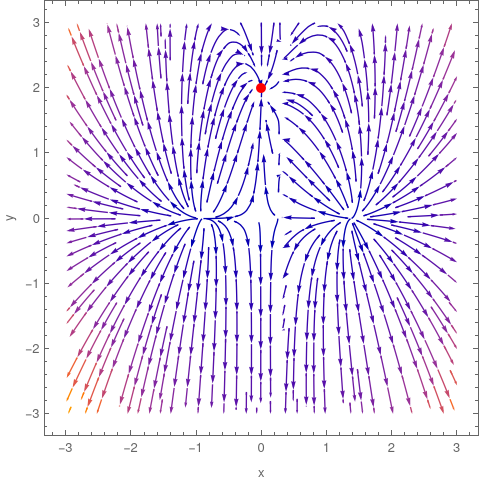}
        \caption{Phase-space portrait of the attractor point $P_2$ for $\mu_0=-1$ and $z=1$.}
        \label{fig:sub1}
    \end{subfigure}
    \hspace{0.02\linewidth}
    \begin{subfigure}[t]{0.31\linewidth}
        \centering
        \includegraphics[width=\linewidth]{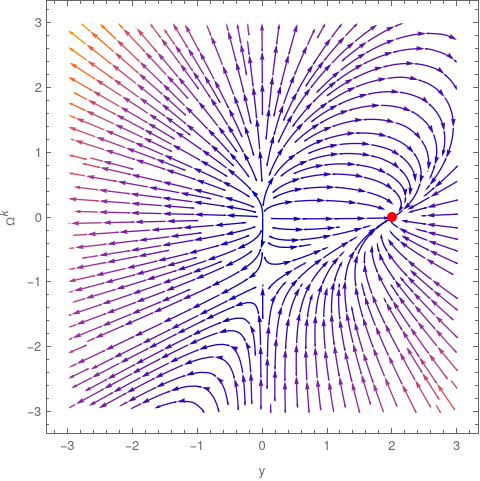}
        \caption{Phase-space portrait of the attractor point $P_2$ for $\mu_0=1$ and $z=1$.}
        \label{fig:sub2}
    \end{subfigure}
    \hspace{0.02\linewidth}
    \begin{subfigure}[t]{0.31\linewidth}
        \centering
        \includegraphics[width=\linewidth]{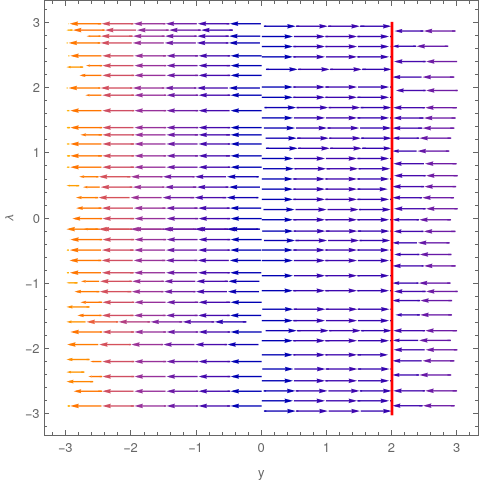}
        \caption{Phase-space portrait showing the vertical attractor line corresponding to $P_2$ for $\mu_0=-1$ and $z=1$.}
        \label{fig:sub3}
    \end{subfigure}
    
    \caption{Phase-space portraits of attractor solution for (\textbf{Case \ref{case III}}).}
    \label{case3_fig1}
\end{figure}

\begin{figure}[htbp]
    \centering
    \includegraphics[width=0.7\linewidth]{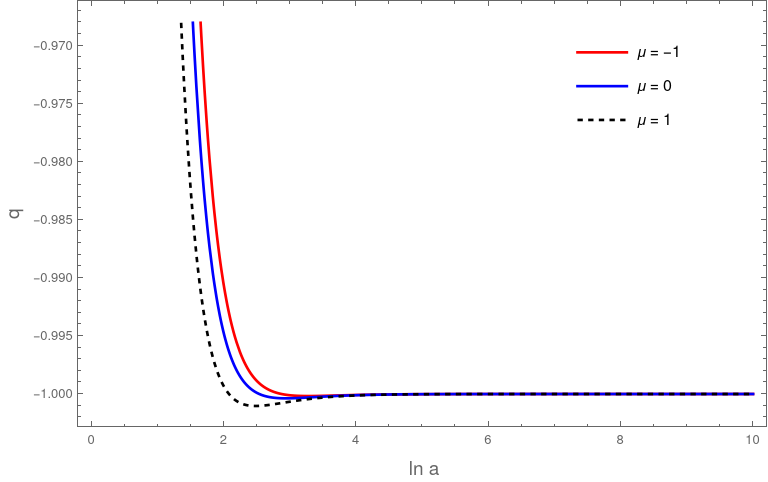}
    \caption{Qualitative evolution of the deceleration parameter of the dynamical system (\ref{case3_xeq})-(\ref{case3_mueq})
for different values of $\mu_0$ with initial conditions ($x[0]=0.1,~y[0]=0.5 ,~z[0]=0.1,~\Omega^{k}[0]=0.1,~\lambda[0]=0$) for
(\textbf{Case \ref{case III}}).}
    \label{case3_fig2}
\end{figure}

The analysis of each fixed point is summarized in Tables \ref{case3table1} and \ref{case3table2}. The stationary points corresponding to a spatially flat universe are $P_1$ to $P_6$ respectively. The fixed point $P_1$ corresponds to a matter-dominated universe with an unstable nature. The critical point $P_2$ represents a de Sitter late-time attractor solution, exhibiting stable behavior. The points $P_3$ and $P_4$ correspond to a decelerating universe and behave as saddle points. The physical properties of $P_5$ and $P_6$ are parametric dependent.  From $P_7$ to $P_{10}$, the critical points are associated with a spatially curved universe. In which $P_7$ with the vanishing deceleration parameter leads to an unphysical point, while $P_{8\pm}$ are also not viable. The fixed point $P_{9+}$ corresponds to a decelerating universe for $\mu_0>0$, whereas $P_{9-}$ describes an accelerating universe for $\mu_0>0$, however, both have a saddle type nature. The stationary point $P_{10}$ shares the same physical characteristics as $P_8$, as discussed earlier in case \ref{case I}. A few 2D phase portraits for the attractor solution are plotted for different parameter values in Fig \ref{case3_fig1}. The qualitative evolution of the deceleration parameter for various values of $\mu_0$ can be analyzed in Fig \ref{case3_fig2}. \footnote{ In comparison, six critical points were identified in \cite{murtazascalar}, same as the present study corresponding to the zero-curvature case. One stationary point corresponded to a matter-dominated universe and another represented a de Sitter attractor, both consistent with our present findings. However, two critical points described stiff-fluid solutions, whereas we find two points characterizing a decelerating universe. Furthermore, two fixed points were parameter-dependent, similar to our points $P_5$ and $P_6$.}

\newpage
\subsection{$\lambda$ and $\mu$ both are variable}\label{case IV}
As discussed earlier, the suitable choices for the coupling function and the potential function are the power law and exponential forms. Based on this, we have four possible cases, as mentioned previously. In this scenario, we find that no new physics arises beyond what has already been analyzed in our earlier discussion. Therefore, we omit this case from the present work and refer the reader to \cite{murtazascalar} for more details.

\section{Concluding remarks}\label{sec3}
In this work, we investigate a non-minimally coupled scalar field theory within the framework of a non-metricity extension of scalar tensor gravity in spatially curved as well as spatially flat spacetime, in a unified way. Our analysis is carried out using the dynamical systems approach, which enables us to reformulate the field equations as a closed system of differential equations by employing a normalized Hubble parametrization. Based on the dynamical system formulation, we find that the relevant scalar field functions are $\mu$ and $\lambda$, which allow us to classify four distinct cases. For each case, we compute the critical points, determine their existence conditions, perform stability analysis, and evaluate the cosmological parameters such as the deceleration parameter $q$ and the effective equation of state parameter $w_{eff}$ in order to extract the cosmological implications associated with each critical point. However, in our analysis, the complete stability characterization of certain critical points is hindered by non-linearities arising from the non-coincident gauge condition. Since the choice of gauge introduces an additional scalar degree of freedom into the system, it necessitates a more specialized stability treatment.

We begin our analysis by considering that both $\mu$ and $\lambda$ are constant, which leads to specific forms of the coupling and potential functions derived from the definitions of our variables. The autonomous system (\ref{case1_eqx})-(\ref{case1_eqOmega}) in this case yields eight critical points, six of them $P_1$ to $P_6$ correspond to a spatially flat universe, while the remaining two $P_7$ and $P_8$ describe a spatially curved universe. The critical point $P_1$ represents a matter-dominated universe with unstable behavior, whereas $P_2$ corresponds to a late-time de Sitter attractor solution. The fixed point $P_4$ describes a decelerated universe with unstable dynamics, while the stationary points $P_3$, $P_5$, and $P_6$ exhibit cosmological properties that depend on the model parameters. The critical point $P_7$ is ruled out due to its vanishing deceleration parameter, whereas the properties of $P_8$ also turn out to be parameter dependent. The phase portraits for an attractor solution in this case are shown in Fig \ref{case1_fig1}, and the quantitative evolution of the deceleration parameter $q$ is also presented in Fig \ref{case1_fig2}.

In the second case, we treat $\lambda$ as a constant parameter while allowing $\mu$ to be a variable. The autonomous dynamical system (\ref{case2_xeq})-(\ref{case2_mueq}) admits six stationary points. Among these, the critical points $P_1$ to $P_3$ correspond to a spatially flat universe, $P_1$ describes an unstable matter-dominated state, $P_2$ represents a stable de Sitter attractor, and $P_3$ characterizes a decelerating universe with unstable behavior. The remaining non-flat critical points are $P_4$ and $P_5$, where $P_4$ is unphysical, while $P_5$, associated with a vanishing deceleration parameter, is also physically unacceptable. The attractor solution $P_2$ for this scenario is illustrated in Fig \ref{case2_fig1}. The qualitative evolution of $q$ for different values of the constant parameter $\lambda_0$ is presented in Fig \ref{case2_fig2}.

In the third scenario, we consider $\mu$ as a constant while treating $\lambda$ as a variable. The corresponding dynamical system yields six fixed points associated with a spatially flat universe. Specifically, 
$P_1$ reflects a matter-dominated universe, while $P_2$ indicates a de Sitter attractor. The critical points $P_3$ and $P_4$ lead to decelerating scenarios, and the properties of $P_5$ and $P_6$ are model parameter dependent. For the non-flat critical points, $P_7$ is not physically acceptable, $P_{\pm8}$ is also not physically viable,  while the physical properties of $P_{\pm9}$ and $P_{10}$ are also parametric dependent.

In the fourth case, where both $\mu$ and $\lambda$ are treated as variables, no new physical insights emerge beyond what has already been discussed in the previous scenarios. Therefore, this case is omitted from our analysis.

Overall, our findings highlight the crucial importance of the non-coincident gauge in non-metricity scalar field theories. This framework not only provides a viable explanation for the universe's late-time acceleration but also guarantees dynamical stability, thereby emphasizing the fundamental role of the non-coincident gauge in cosmic evolution. Further cosmological data analysis will be essential to substantiate the influence of this additional degree of freedom in shaping the history of the universe.
Finally, we should remark here that our conclusions are restricted to background dynamical stability (phase-space stability of critical points) and that perturbative stability is a separate and necessary step for a complete viability assessment. Encouragingly, a recent related work \cite{ganesh} has already developed the scalar cosmological perturbation framework in the scalar–tensor extension of non-metricity gravity, deriving the perturbed field equations and studying the density contrast evolution under the quasi-static approximation. This provides a timely overview and strong foundation for the perturbative sector relevant to our framework.

\section*{Data Availability Statement}
Data sharing is not applicable to this article, as no new datasets were generated or analyzed. The study is entirely theoretical in nature.

\end{document}